\documentclass{article} 
\usepackage{cite,epsfig}
\author{Torsten Henning$^a$\thanks{present address: 
Applied Solid State Physics, G\"oteborg University and Chalmers
University of Technology AB, S-41296 G\"oteborg, Sweden,
e-mail \texttt{henning\symbol{64}fy.chalmers.se}}, 
H Kliem$^a$\thanks{present address: Institute of Electrical Engineering
Physics, 
University of Saarbr\"ucken, D-66041 Saarbr\"ucken, Germany%
}, A Weyers$^b$\thanks{present address: Geod\"atisches Institut, RWTH,
D-52062 Aachen, Germany} and W Bauhofer$^a$\\
\small $^a$TU Hamburg-Harburg, AB Materialien der Mikroelektronik,\\[-0.8ex]
\small D-21071 Hamburg, Germany\\[-0.5ex]
\small $^b$Institut f\"ur Werkstoffe der Elektrotechnik,
\small  RWTH,  D-52056 Aachen, Germany}
\title{Characterisation of HTSC ceramics from their resistive transition}
\date{cond-mat/9707165 (1997-07-16)\\
to be published in Supercond. Sci. Technol.}
\begin{document}
\bibliographystyle{unsrt}
\maketitle
\frenchspacing 
\begin{abstract} 
The resistivity vs.{} temperature relation in bulk ceramic HTSC under
self-field conditions as well as
in weak external magnetic fields is modelled
by local Lorentz force induced fluxon motion with
temperature dependent pinning.
A pinning force density and two viscous drag coefficients in intergrain
and intragrain regions, respectively, 
can be used as characteristic
parameters describing the temperature, current, and 
external field dependences
of the sample resistance.
\end{abstract}

\section{Introduction} 

The resistivity of a ceramic superconductor is known to depend on
transport current,
temperature,  external magnetic field, and on the
geometry of the sample 
\cite{palstra:89:apl}.  The critical
current also depends on the latter three parameters,
which means that the averaged critical current density alone
is unsuitable as a
parameter for material characterisation.

Resistance in cuprate superconductors is supposed to be caused by
fluxons \cite{brandt:95:fll}
moving under the influence of Lorentz forces and pinning forces
acting in opposite directions, leading to a superconducting state with
finite current density and truly zero resistance.

In earlier papers 
\cite{kliem:91:self,weyers:92:crit}, a model for fluxon
motion in HTSC was introduced that contains temperature dependence
through the temperature dependence of pinning forces and fluxon
mobilities, and in which a $B^{-1}$ magnetic field dependence of the
pinning forces causes the critical currents to decline rapidly with
increasing fields.  This Lorentz force model, 
which is an alternative to the
model of thermally activated fluxon motion, has already explained the
temperature and magnetic field dependence of local critical current
density 
\cite{schatteburg:94:jap,kilic:95:selffield,kliem:96:sst}
as well as its variation with sample cross section.

We report  measurements of the temperature dependent
resistivity and the broadening of the resistive transition of YBaCuO
with increasing transport current or external magnetic field in the
temperature range (77\dots 91)\,K.  We
demonstrate how the broadening can be described in terms of 
Lorentz force induced fluxon motion.

The resistivity of ceramics shows a behaviour that can
be described with a model based on Lorentz force induced fluxon motion
in two phases. Numerical model calculations and their comparison with
experimental data show that three sample parameters are sufficient to
describe the $\varrho(T;I,B_{\mbox{\it\footnotesize ext}})$
characteristics. 

\section{Experiments}

\subsection{Technique} 

To avoid errors in temperature dependent resistivity measurements by
ohmic heating of the contacts or the bulk we employed a transient
current signal technique, using a single triangular shaped current
pulse. Choosing equal rise and fall rates and averaging the measured
voltage drop along the superconductor for rising and falling current
compensated the voltage background due to induction.

The voltage $U$ was amplified by a calibrated differential amplifier
switchable over several decades and registered on a two channel
transient recorder together with the measured current $I$. From these
raw data, $\varrho(T,I)$ was calculated for five fixed $I$ values per
sweep. Duration of the current triangle was $1/1800\;\mbox{s}$, and we
evaluated $U$ for $I<10\;\mbox{A}$.  Experiments with transient signals
were performed at temperatures above $77\,\mbox{K}$, and the temperature
was monitored by a NiCr-CuNi thermocouple.  Static magnetic fields of
several millitesla could be applied perpendicular to the direction of
the current flow.

YBaCuO samples of typical cross sections $2.5\times2.5\,\mbox{mm}^2$ or
$4\times8\,\mbox{mm}^2$ and length $20\,\mbox{mm}$ had been prepared by
the standard mixed oxide method. We made current leads by soldering on a
$1.5\,\mu\mbox{m}$ copper layer deposited on the ceramic by thermal
evaporation. Contact resistivities were
$0.5\dots4\,\mbox{m$\Omega\,$cm}^2$ at room temperature, and we found
that the mechanical stability of the 
soldering joint was sufficient as long as  the
contact resistivities did not exceed these values.

\subsection{Measurements} 

\begin{figure}
\begin{center}
\epsfig{file=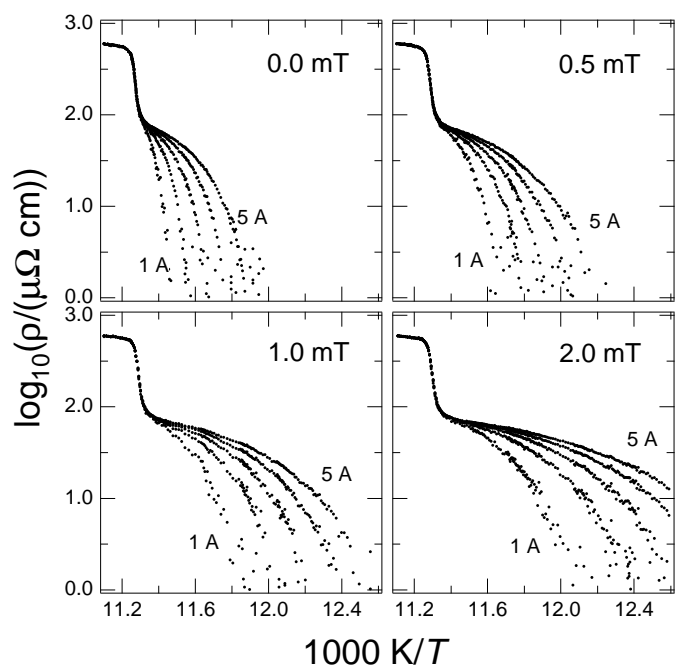,width=0.75\textwidth}
\end{center}
\caption{\label{cordula15oct}Resistive transition of YBaCuO sample V (see
  Table~\ref{tabelle}) for transport currents $I=(1,\dots,5)\,\mbox{A}$ in
  different external magnetic fields.}
\end{figure}

Figure~\ref{cordula15oct} shows 
a set of Arrhenius plots of the resistive
transition of our YBaCuO sample V for five different transport currents
in zero and three different external magnetic fields. The most
remarkable feature of these curves, apart from the overall broadening of
the transition with both parameters, is their inclination at
(for this sample) about a tenth of the resistivity value at the onset of
metallic behaviour. We refer to the resistivity value at this point as
$\varrho_{\mbox{\it\footnotesize incl}}$. The inclination point
separates the broad shoulder of the curves \cite{abele:93:xxx} from a
leap in resistivity over a comparatively small temperature interval.

For the range of transport currents and magnetic fields in the situation
of fig.~\ref{cordula15oct}, the broadening of the transition occurs only
in the shoulder, while the leap is not affected. The inclination at
$\varrho_{\mbox{\it\footnotesize incl}}$ thus becomes sharper. The
critical state, i.\,e. the point where the curves diverge to infinity,
is shifted towards lower temperatures with increasing field or transport
current.

\begin{figure}
\begin{center}
\epsfig{file=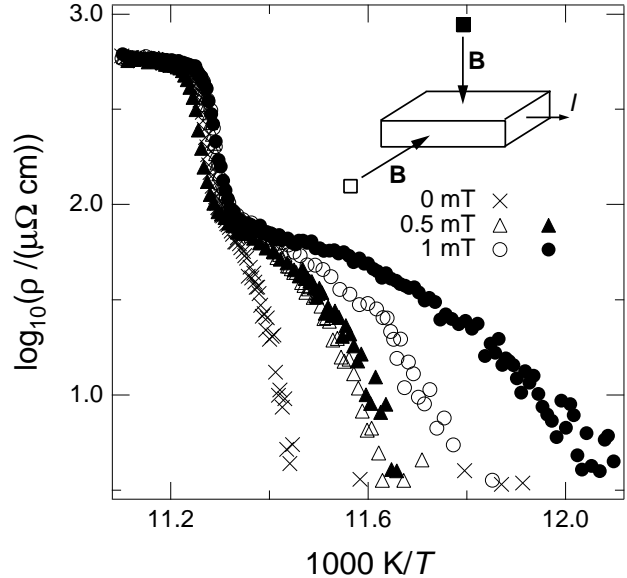,width=0.75\textwidth}
\end{center}
\caption{\label{cordulaorientierung}Geometry (orientation) 
  dependence of the resistive transition of sample V in
  external magnetic fields. Filled symbols indicate that the field was
  applied perpendicular to the larger lateral face of the sample.}
\end{figure}

It is worth noting that the width of the transition depends on the
orientation of the sample to the magnetic
field. Fig.~\ref{cordulaorientierung} gives a set of experimental
data for sample V, showing that the transition is wider when the field
is applied perpendicular to the larger lateral face of the sample.

\begin{figure}
\begin{center}
\epsfig{file=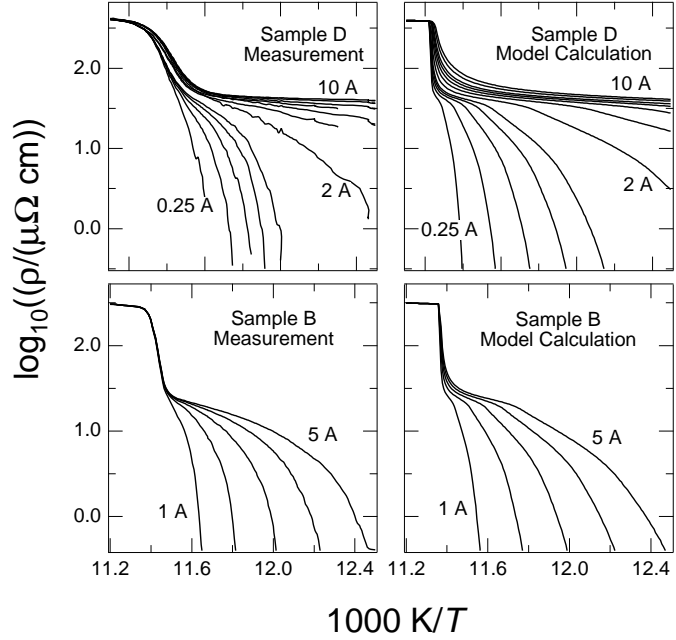,width=0.75\textwidth}
\end{center}
\caption{\label{eigenfeldvergleich}Comparison of the
  measured resistive transitions (left) and 
  model calculations in the two phase model for
  YBaCuO samples D (top) and B (bottom)
  under self field conditions. See table~\ref{tabelle} for
  parameters.}
\end{figure}

\begin{figure}
\begin{center}
\epsfig{file=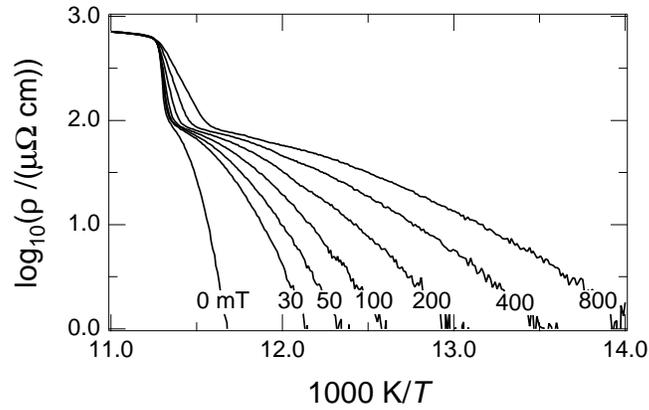,width=0.75\textwidth}
\end{center}
\caption{\label{cordulamittelfeld}Broadening of the resistive transition 
  of YBaCuO sample V
  in external magnetic fields, measured by dc technique with $I=100$\,mA.}
\end{figure}

The transition above $\varrho_{\mbox{\it\footnotesize incl}}$ is also
broadened with transport currents and magnetic fields, though 
much less than the shoulder. For the case of current dependent
broadening, we can see this in the top left panel of
fig.~\ref{eigenfeldvergleich}, where data for sample D are shown. This
sample had a smaller cross section than sample V, and we could reach
higher current densities in the experiment.
Broadening with external magnetic fields is obvious from
fig.~\ref{cordulamittelfeld}, which shows the results of a resistance
measurement with standard dc technique in external magnetic fields up to
800\,mT.

The position of the inclination point $\varrho_{\mbox{\it\footnotesize
incl}}$, however, is not affected by the broadening of the resistivity
above $\varrho_{\mbox{\it\footnotesize incl}}$. 
We interpret this as evidence for a two phase behaviour of the
resistive transition, where both phases differ in the parameters
describing the pinning.

\section{Model assumptions}

\subsection{Lorentz force model} 

The model of Lorentz force induced fluxon motion causing resistance in
HTSC has been described in more detail in earlier papers
\cite{kliem:91:self,weyers:92:crit},
and applied to describe experiments on the critical current density
of HTSC
\cite{kilic:95:selffield,kliem:96:sst} and the redidtribution of local
current densities in external fields \cite{schatteburg:94:jap}. 
In this section, we
briefly review the foundations here as they were stated and used in
these references.\par  
It is assumed that a flux line moves
with a velocity $v$ determined by a viscous drag coefficient $\eta$ and
the excess of local Lorentz force density (per unit length) $F_L=\Phi_0
j$ over the pinning force density $F_p$,
\begin{equation}
v=\cases{\frac{\displaystyle
			1}{\displaystyle\eta(T)}\left(F_L(j)-F_p(B,T)\right)
			& if $F_L\ge{F_p}_1$,\cr \phantom{a} & \cr 0 & if
			$F_L<{F_p}_1$.\cr}
\end{equation}
For $F_L={F_p}_1$, the sample is in the critical state. The pinning
forces $F_p$ are assumed to be distributed evenly over an interval
$[{F_p}_1,{F_p}_1+\Delta F_p]$ \cite{kliem:91:self}.
This results in an average fluxon velocity \cite{kliem:91:self}
\begin{equation}
\langle v\rangle=\int_0^{F_p(j)}\,v\left(F_p^\prime\right)\,%
\frac{\displaystyle 1}%
{\displaystyle \Delta{}F_p}\,dF_p^\prime.
\end{equation}
The motion of vortices with area density $n$ causes losses leading to an
electric field \cite{roseinnes:78:xxx}
\begin{equation}
E=n\Phi_0 \langle v\rangle .
\end{equation}
Pinning forces depend inversely on the local field of magnetic induction,
\begin{equation}
F_p=\frac{\displaystyle \Phi_0 F_0(T)}{B},\qquad
F_{p1}=\frac{\displaystyle \Phi_0 F_{01}(T)}{B},\qquad
\Delta{}F_p=\frac{\displaystyle \Phi_0 \Delta{}F_0(T)}{B}.
\end{equation}
This relation was used both in the critical current studies of
Kilic \cite{kilic:95:selffield} and Kliem et al. \cite{kliem:96:sst} and
should be applicable here since the applied external magnetic
fields were of the same order of magnitude as the self-field,
like in the  paper on current densities in small external fields
by Schatteburg et al. \cite{schatteburg:94:jap}.
$B$ is the sum of self-field and external field, for the latter we
neglect magnetisation effects.  As a consequence of the
above assumptions,
we get
\begin{equation}
\label{ecaseseqn}
E=\cases{\frac{\displaystyle
\Phi_0}{\displaystyle 2\eta\,\Delta F_0}\left(j\cdot B-F_{01}\right)^2
& if ${F_p}_1\le F_L\le {F_p}_1+\Delta F_p$,\cr \phantom{a} & \cr 
\frac{\displaystyle \Phi_0}{\displaystyle \eta}\left(
j\cdot B-\frac{1}{2}\left(2 F_{01}+\Delta F_0\right)\right)& if
$F_L\ge {F_p}_1+\Delta F_p$.\cr}
\end{equation}
Since the electric field is constant throughout the cross section,
$j$ as well as $B$ depend on $r$, and (\ref{ecaseseqn}) is a local
relation 
\begin{equation}
\label{localeqn}
j(r)\,B(r)=\alpha,
\end{equation}
which has been found to agree in the critical state with experimental
results on spatial current distribution \cite{schatteburg:94:jap}
and ac permeability measurements \cite{gjoelmesli:93:xxx}.
With (\ref{ecaseseqn}) and (\ref{localeqn}) and Maxwell's equation
\begin{equation}
\frac{\displaystyle 1}{\displaystyle \mu_0} \oint_C B\,dl=
\int_A j\,dA,
\end{equation}
$j(r)$, $B(r)$ and $E$ can be calculated
\cite{weyers:92:crit,schatteburg:94:jap,kliem:96:sst}.
The pinning force
density and the viscous drag coefficient depend on temperature via the
upper critical field $H_{c2}$, for which we assume
\begin{equation}
H_{c2}=H_{c2}(0)\cdot\left(1-\left(\frac{T}{T_c}\right)^2\right).
\end{equation}
The temperature dependencies $F_0(T)$ and $\eta(T)$
are entirely
responsible for the temperature dependence of the resistivity
$\varrho(T)$.
For ceramics, Kliem et al. found from measurements of the temperature
dependence of the critical current \cite{kliem:96:sst} that
\begin{equation}
F_0(T)=\tilde{F}_0\cdot\left(1-\left(\frac{T}{T_c}\right)^2\right)^2
\end{equation}
gave the best fit to the experimental data. Using 
a similar model, Kilic
\cite{kilic:95:selffield} found that the last exponent should be 1.5 in
the case of thin films. For the temperature dependence of the viscous
drag coefficient, we use 
\begin{equation}
\eta(T)=\frac{\displaystyle H_{c2}(T)}%
{\displaystyle\varrho^{\mbox{\footnotesize\rm v}}}=
\tilde{\eta}\cdot\left(1-\left(\frac{T}{T_c}\right)^2\right)
\end{equation}
from the same reference \cite{kilic:95:selffield}.
Both $\varrho^{\mbox{\footnotesize\rm
v}}$ and $\tilde{\eta}$ are  independent of temperature.

The fluxon radius increases with temperature  and
the fluxon density increases with magnetic field strength. This can
result in an overlap of normal conducting flux cores. Simultaneously the
fluxon velocity and therefore the resistivity increase with temperature
and magnetic field. Instabilities resulting from this are eliminated by
the following assumption, which turns out to fit well to experiments:
those parts of the sample's cross section in which the resistivity
caused by fluxon motion would exceed the normal state resistivity,
become normal conducting.
For a suitable
choice of parameters, the resistivity curves in an Arrhenius plot fan
out from the onset of metallic conductivity and are in segments
linear. To reduce the number of parameters in our model calculations, we
have assumed $\Delta F_p=F_{p1}$ for the following. These are the
assumptions that constitute what we refer to as the single phase model.

\subsection{Two phase model} 

Taking into account the polycrystalline nature of sintered HTSC
ceramics, we assume that the current flows through intergrain and
intragrain regions and that these phases differ in fluxon mobilities. A
relatively narrow transition in the intragrain phase and a broad
transition in the intergrain phase lead to the observed 
leap-and-shoulder
structure of the resistivity curves when we treat both phases as
resistors connected in series, i.\,e., carrying the same
current. Each of the phases is then treated  separately according 
to the single phase model assumptions
given above.\par
This is the simplest possible extension of the Lorentz force model to
two phase ceramics. In this phenomenological approach, we do not make
assumptions on what the grain boundaries look like or how grains are
connected. 

\section{Comparison with experiment}

\begin{table} 
\caption{\label{tabelle}Characteristic parameters of 
three YBaCuO samples obtained from comparison of model calculations to the
measured resistive transitions under self-field conditions.}
\vskip1.5ex
\begin{tabular}{ll||r|r|r}
Quantity & Unit & Sample D & Sample B & Sample V\\ \hline
\vphantom{{\large (}}cross section & mm$^2$ & $2.4\times2.4$ &
$2.5\times2.5$ & $4\times8$ \\ $j_c$ at $77\,\mbox{K}$ & A$\,$cm$^{-2}$
& 32 & 38 & 36 \\ \hline \vphantom{{\large (}}$T_{\mbox{\it onset}}$ & K
& 86 & 88 & 88 \\ intergrain fraction & &1:9 & 1:13 & 1:13 \\ \hline
\vphantom{{\large (}}$F_{p1}$ at $77\,\mbox{K}$ & Nm$^{-3}$ & 50 & 600 &
200 \\ $\tilde{\eta}$ (intergrain) &
$10^{-12}\,\mbox{kg}\,\mbox{m}^{-1}\,\mbox{s}^{-1}$ & 15 & 6 & 6 \\
$\tilde{\eta}$ (intragrain) &
$10^{-12}\,\mbox{kg}\,\mbox{m}^{-1}\,\mbox{s}^{-1}$ & 1500 & 600 & 600
\\
\end{tabular}
\end{table}

\subsection{Self-field conditions} 

An analytical solution for $I(E;T)$ exists for cylindrical cross section
and self-field conditions. We have performed measurements on samples
with square cross section and treated them for the model calculations as
cylinders with the same cross section area.

\begin{figure}
\begin{center}
\epsfig{file=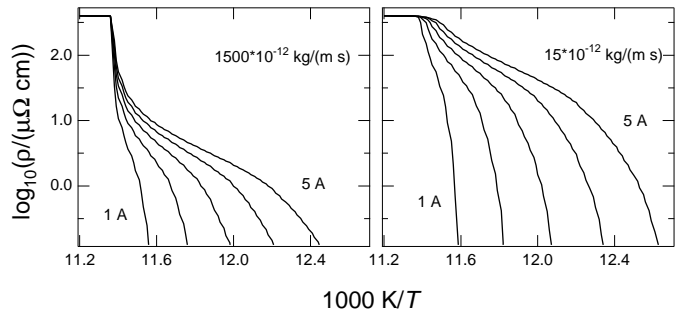,width=0.75\textwidth}
\end{center}
\caption{\label{einphassim}Single phase model calculation:
  effect of the viscous drag coefficient 
  $\tilde{\eta}$ on the
  resistivity vs.\ temperature (Arrhenius plot) curves
  under self field conditions.}
\end{figure}

The viscous drag coefficient $\tilde{\eta}$ determines the
shape of the transition. Fig.~\ref{einphassim} shows calculated
resistive transitions in the single phase model, with parameters chosen
to model the behaviour of sample D. We see how the curvature changes
with increasing $\tilde{\eta}$ and how the curves in the Arrhenius plot
start sagging, but the sharp inclination found in the experiment could
not be rendered in single phase model calculations.

Fig.~\ref{eigenfeldvergleich} shows
a comparison of measurements and model
calculations under self field conditions
for two samples. Taking for $T_c$ the onset temperature of
metallic conductivity and for the fraction of the intergrain phase in
the sample the ratio $\varrho_{\mbox{\it\footnotesize incl}}/%
\varrho_{\mbox{\it\footnotesize onset}}$, we have varied $F_1$ and
$\tilde{\eta}$ in the intergrain and intragrain phase to fit the whole
set of curves. The parameters are given in Table~\ref{tabelle}.

While the shoulder and the inclination point of the curves are modelled
quite well, the roundness near the onset of metallic conductivity is not
given correctly by the calculation. This is probably not caused simply by
fluxon motion, but has physically different origin \cite{hikita:90:fluct}.

\subsection{External magnetic fields} 

For a rectangular sample cross section and in external magnetic fields,
the model calculation of $E(I;T,B)$ was performed numerically by a
self-consistent iterative computation of current density and magnetic
field on an $n\times m$ grid.  We typically used $n$ and $m$ values of
$10\dots20$ and started from a current distribution with constant
density.
\begin{figure}
\begin{center}
\epsfig{file=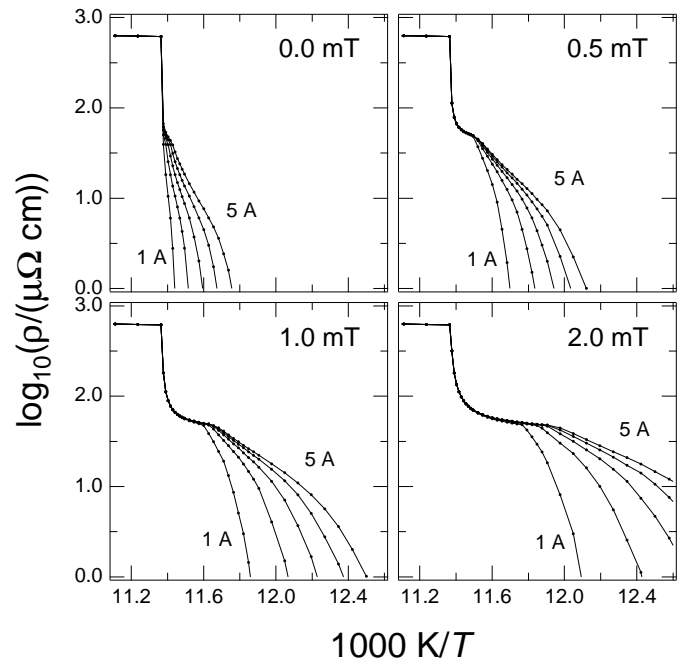,width=0.75\textwidth}
\end{center}
\caption{\label{cordulasimulation}Numerical simulation 
  in the two phase model for 
  the resistive transition of YBaCuO sample
  V for different transport currents and external
  magnetic fields. Compare with fig.~\ref{cordula15oct};
  parameters see Table~\ref{tabelle}.}
\end{figure}
Fig.~\ref{cordulasimulation} shows the results of a model calculation
for sample V
with the parameter set (see Table~\protect\ref{tabelle})
determined from the measurements and calculations under self-field
conditions. Comparison of the calculations in
fig.~\ref{cordulasimulation} with the experimental data in
fig.~\ref{cordula15oct} shows that the behaviour in external fields is
reproduced without adding any further parameters to the model.

The agreement of the model calculations with the experiment is good, though
for zero external  field, the measured transition broadening is slightly
larger than calculated. The model calculations, however, do not render
the observed dependence from magnetic field orientation for rectangular
cross section.  This dependence might be caused by shielding effects
leading to different macroscopic demagnetisation factors for both
orientations.

A deviation from the model predictions for small external fields
attributed to shielding effects was also found in measurements of the
spatial distribution of the critical current \cite{schatteburg:94:jap}.

\section{Conclusion} 

We have demonstrated that 
a previously developed phenomenological
model based on the concept of Lorentz force induced fluxon
motion can be applied to the resistive transition of YBaCuO under
self-field conditions as well as in (small) external fields. For very
small fields, slight quantitative deviations from the model predictions
occur which we attribute to shielding effects and which deserve further
investigation.  A simple
two phase model allows the characterisation of YBaCuO
ceramics by a pinning parameter, two fluxon mobilities, the onset
temperature of metallic conductivity and the intergranular phase
fraction. These parameters can be determined by comparison of the
current dependent broadening of the resistive transition,
obtainable in relatively simple transport measurements, to numerical
model calculations. Even samples with similar onset temperature
and critical current density can differ considerably in their resistive
transition above the critical state, and our results show a way of
describing and quantifying these different behaviours
independent of the individual sample geometry.
\section*{Acknowledgements} 
We thank Dr. Knechtel (TU Hamburg-Harburg, AB Technische Keramik) 
for the preparation of YBaCuO samples.
\bibliography{p97}
\end{document}